\documentclass{article}
\newcommand{\bfr}{\begin{flushright}}
\newcommand{\efr}{\end{flushright}}
 
\begin{document}
\title{Bose-Einstein Condensation in Compactified Spaces
}
\author{Kiyoshi Shiraishi\\
Department of Physics, Tokyo Metropolitan University,
Tokyo 158
}
\date{Prog. Theor. Phys. {\bf 77} (1987) pp. 975--982
}
\maketitle
\begin{abstract}
We discuss the thermodynamic potential of a charged Bose gas with the
chemical potential in arbitrary dimensions. The critical temperature
for Bose-Einstein condensation is investigated. In the case of the
compactified background metric, it is shown that the critical
temperature depends on the size of the extra spaces. The asymmetry of
the ``Kaluza-Klein charge'' is also discussed.
\end{abstract}

\section{Introduction}
Recent progress in the unification of gauge interactions has been
made in higher dimensions. The original idea of Kaluza and
Klein \cite{1} is reviving after about a half century.

In the framework of the so-called Kaluza-Klein theories, gauge fields
and symmetries are interpreted as the part of the gravitational field
and the symmetries of extra spaces respectively.\cite{2} The revival of
Kaluza-Klein theories is activated by the development of supergravity
theories.\cite{3} However, many difficulties have been pointed out:
such as the chiral fermion problem, unfavorable quantum nature. In
many cases, it is well known that the introduction of primary gauge
fields can solve the problems.\cite{4} As a result of recent progress
in ten-dimensional superstring theories,\cite{5} it is suggested that
the gauge fields may be contained in the higher-dimensional theories,

On the other hand, the source of the gauge field, that is, the
charge, is a conserved quantity if there is no symmetry-breaking. In
our universe, there is no reason why the certain charge asymmetry
should vanish from the beginning (of the universe). In the presence of
the charge asymmetry, a Fermi gas becomes strongly degenerate at low
temperatures. For a Bose gas, under the same circumstances, it is
expected that Bose-Einstein condensation takes place. The effect of
the charge-asymmetry may also play an important role in the phase
transitions in very early universe.\cite{6}

In our previous paper,\cite{7} the finite density effects for Dirac
femion fields in higher dimensions are discussed. The
compactification in Kaluza-Klein theories may be also influenced by the
behavior of the bosonic matter fields. Especially in supersymmetric
theories, the contribution of bosonic fields as well as fermionic
matter fields is important.

In the present paper, we consider the properties of a charged Bose gas
with  non-zero chemical potential and derive the expressions for the
critical temperature for Bose-Einstein condensation in arbitrary
dimensions and in ``Kaluza-Klein background'' geometries. The general
expression for the thermodynamic potential of a charged scalar boson
field is obtained in \S 2. We also explain how we can determine the
critical temperature for condensation in arbitrary flat space. The
generalization to
(partly) compact spaces is made in \S 3. The charge stated above is the
primary one in the higher dimensions. We can consider the charge
induced by the compactification, i.e., by the Kaluza-Klein scheme, and
its excess in the universe. In \S 4, we discuss them, and in particular,
investigate whether condensation occurs or not. Finally, \S 5 is
devoted to discussion.

\section{Critical temperature in flat arbitrary dimensions}
Recently, several authors discussed the expressions of the
thermodynamic potential $\Omega$ for a Bose gas with mass $M$ and
chemical potential $\mu$ in flat $d$-dimensional space at temperature
$T=\beta^{-1}$.\cite{8,9} In this section, we would like to get them by
another way which was used for a fermion field in the previous
paper.\cite{7}

Allen\cite{8} showed that the thermal partition function can be
derived from one-loop effect in a quantum field theory with
imaginary time.\cite{11} It was shown that the vacuum energy which
needs to be regularized is distinguished from others. Granted that the
vacuum energy can be discarded, we find the thermodynamic potential for
a charged scalar boson field by using the heat-kernel(-like)
method:\cite{7}
{\small
\begin{eqnarray}
\Omega&=&-\frac{V_d}{(4\pi)^{1/2}}
\int_0^\infty dt\,t^{-3/2}\int\frac{d^d{\bf k}}{(2\pi)^d}2
\sum_{n=1}^\infty\cosh(n\beta\mu)\exp\left\{-t({\bf k}^2+M^2)-
\frac{\beta^2n^2}{4t}\right\}\nonumber \\
&=&
-\frac{V_d}{(4\pi)^{(d+1)/2}}
\int_0^\infty dt\,t^{-(d+1)/2-1}2
\sum_{n=1}^\infty\cosh(n\beta\mu)\exp\left\{-t M^2-
\frac{\beta^2n^2}{4t}\right\}\nonumber \\
&=&
-2\frac{V_d}{(4\pi)^{(d+1)/2}}\beta^{-(d+1)}
\sum_{n=1}^\infty\cosh(n\beta\mu)2\left(\frac{2\beta
M}{n}\right)^{(d+1)/2}K_{(d+1)/2}(\beta Mn)\,,
\label{2.1}
\end{eqnarray}
}
where $V_d$ denotes the volume of $d$-dimensional space.
(Compare with (2.10) in Ref. \cite{7}).)

The first line of (\ref{2.1}) was originated from the zeta function
regularization method.\cite{7} This form of the expression will turn out
to be useful in the later section.

In order to evaluate the thermodynamic potential as for this
expression $\Omega$, we must make the expansion for $\Omega$ in terms of
infinite series, i.e., high-temperature expansions. Nevertheless, naive
procedure of expansion and resummation is liable to miss the term
which has come from the branch point singularity at $\mu=M$ in the
complex $\mu$-plane. For instance, the thermodynamic potential for a
charged scalar field in even $d$-dimensional space contains the
following form:\cite{8}
\begin{equation}
\Delta\Omega=(-1)^{(d-1)/2}\frac{\pi V_d
T}{(4\pi)^{d/2}\Gamma((d+2)/2)}(M^2-\mu^2)^{d/2}\,.    
\label{2.2} 
\end{equation}
Thus, we should be careful when we calculate the
high-temperature expansions of some thermodynamical quantities. In
the present paper, we pay attention only to Bose-Einstein
condensation and the critical temperature for it, and one can avoid
here the complicated issue as one will see later.

In order to examine the property of Bose-Einstein condensations, we
show the charge density of the system in the familiar form as
follows:
{\small
\begin{eqnarray}
\rho&=&-\frac{1}{V_d}\frac{\partial\Omega}{\partial\mu}\nonumber \\
&=&
\frac{2}{(4\pi)^{(d+1)/2}}\beta^{-d}
\sum_{n=1}^\infty n\sinh(n\beta\mu)2\left(\frac{2\beta
M}{n}\right)^{(d+1)/2}K_{(d+1)/2}(\beta Mn)\nonumber \\
&=&
\frac{2M^d}{(4\pi)^{d/2}\Gamma(d/2)}
\int_1^\infty
dx\,(x^2-1)^{(d/2)-1}\left\{\frac{1}{e^{\beta(Mx-\mu)}-1}-
(\mu\rightarrow
-\mu)\right\}\,,
\label{2.3}
\end{eqnarray}
}
for $d>0$.
 At the step into the third line, we used the integral representation
\begin{equation}
K_\nu(z)=\frac{\sqrt{\pi}(z/2)^\nu}{\Gamma(\nu+1/2)}\int_1^\infty
e^{-zx}(x^2-1)^{\nu-1/2}dx\,.
\label{2.4}
\end{equation}
It is easy to see that the final form of (\ref{2.3}) is equivalent to
\begin{equation}
\rho=\int_0^\infty\frac{d^d{\bf k}}{(2\pi)^d}
\left[\frac{1}{\exp\{\beta(\sqrt{{\bf
k}^2+M^2}-\mu)\}-1}-(\mu\rightarrow-\mu)\right]\,.
\label{2.5}
\end{equation}

Here, we briefly review the argument for the critical
temperature.\cite{9} Above some critical temperature $T_c$, we can
always find
$\mu<M$ satisfying (\ref{2.5}) if $\rho$ fixed. Below $T_c$, no such
$\mu$ can be chosen and we fail in reading (\ref{2.4}) as it is.
Equation (\ref{2.5}) must be interpreted as the charge density in the
${\bf k}\ne 0$ state. In other words, if (\ref{2.5}) remains finite in
the limit
$\mu\rightarrow M$, the rest of the charge must condensate at the
${\bf k}=0$ state. Therefore, we have only to know the behavior of the
expression of the charge density in the limit
$\mu\rightarrow M$ in order to know the critical value for the
condensation; furthermore, we can keep away from the complication of
$\Delta\Omega$ in (\ref{2.2}) at least in flat
$d$ dimensions.

We observe that the charge of the ${\bf k}\ne 0$ state expressed by
(\ref{2.3}) and (\ref{2.5}) is finite even in the limit $\mu\rightarrow
M$ provided
$d>2$. One can find the critical value $T_c$ satisfies the following
relation at
$\mu=M$ in the high-temperature limit:
\begin{equation}
\rho=\frac{2}{\pi^{(d+1)/2}}\Gamma\left(\frac{d+1}{2}\right)
T_c^{d-1}M\,\zeta(d-1)\, ,        
\label{2.6}
\end{equation}
where $\zeta(z)$ is Riemann's zeta function. Hence, we find
\begin{equation}
T_c=\left[\frac{\pi^{(d+1)/2}\rho}{2\Gamma((d+1)/2)\zeta(d-1)M
}\right]^{1/(d-1)}\,,     
\label{2.7)}
\end{equation}
and below $T_c$, ${\bf k}=0$ states cannot carry the charge of scalar
particles, so we conclude that there occurs condensation. The physical
interpretation of condensation, especially for massless particles, can
be found in Ref.~\cite{4}. So far we gave an overview of Bose-Einstein
condensation in flat space, but around here, we shall turn to discuss
condensation in compact spaces.

\section{Partly compactified space}
In this section, we treat the scalar fields in the background geometry
$T \times R^d \times S^N$, which has been discussed frequently as the
model space of non-abelian Kaluza-Klein
theories. As in the fermion case,\cite{7} we can write down the
expressions of thermodynamic potential for the scalar boson as
follows:
\begin{eqnarray}
\Omega&=&
-\frac{V_d}{(4\pi)^{(d+1)/2}}
\int_0^\infty dt\,t^{-(d+1)/2-1}2
\sum_{n=1}^\infty\cosh(n\beta\mu)\nonumber \\
&&\times\sum_{\ell=0}^\infty
d_\ell\,\exp\left\{-t (\omega_\ell^2+M^2)-
\frac{\beta^2n^2}{4t}\right\}\nonumber \\
&=&
-\frac{2V_d}{(4\pi)^{(d+1)/2}}T^{d+1}
\sum_{n=1}^\infty\cosh(n\beta\mu)\sum_\ell d_\ell 2\left(\frac{2\beta
\sqrt{\omega_\ell^2+M^2}}{n}\right)^{(d+1)/2}\nonumber
\\
&&\times K_{(d+1)/2}(\beta
\sqrt{\omega_\ell^2+M^2}n)\,,
\label{3.1}
\end{eqnarray}
where the degeneracy
$d_\ell=\{(2\ell+N-1)\Gamma(\ell+N-1)\}/\{\ell!\Gamma(N)\}$, and
$\omega_\ell^2=\ell(\ell+N-1)/a^2$; $a$ is the scale of $S^N$. The
chemical potential
$\mu$ has been introduced for an elementary charge of the complex scalar
field in
$(1+d+N)$ dimensions here. Therefore the total charge density of this
system is
\begin{equation}
\rho=\sum_\ell d_\ell\int_0^\infty\frac{d^d{\bf k}}{(2\pi)^d}
\left[\frac{1}{\exp\{\beta(\sqrt{{\bf
k}^2+M^2+\omega_\ell^2}-\mu)\}-1}-(\mu\rightarrow-\mu)\right]\,.
\label{3.2}
\end{equation}
Here, we used the representation of the modified Bessel function
(\ref{2.4}) again. The above expressions show that the thermodynanic
quantities such as $\rho$ are sum of the ones in $R^d$ for the scalar
particles with mass
$\sqrt{M^2+\omega_\ell^2}$. In other words, this allows to interpret for
the thermodynamical quantities being merely the sum of ones for each
``pyrgon''\cite{12} state up to degeneracy, in general.

Let us investigate condensation in the space $R^d \times S^N$. To derive
the critical value for $T$, we must know the value of $\mu$ when
condensation occurs. In a curved space, setting $\mu=M$ is not so
trivial. However, a physical interpretation in this case is easily
obtained to conclude that only pyrgon state which has lowest mass is
condensate. (We call even zero mass level a ``pyrgon state'' in this
paper.) Thus, we set
$\mu=M$ at the critical point.

If $T\ll M$, all pyrgon states can be treated as non-relativistic.
Namely, in (\ref{3.1}) we can use the asymptotic fom1 of the Bessel
function:
\begin{equation}
K_\nu(z)\stackrel{z\gg 1}{\longrightarrow}
\sqrt{\frac{\pi}{2z}}e^{-z}\,.
\label{3.3}
\end{equation}
Using this, we can get
{\footnotesize
\begin{equation}
\rho\sim\sum_\ell d_\ell \frac{1}{(2\pi)^{d/2}}
T_c^{d/2}(\sqrt{\omega_\ell^2+M^2})^{d/2}
\sum_{n=1}^\infty
\frac{1}{n^{d/2}}
\exp\{-n\beta(\sqrt{\omega_\ell^2+M^2}-M)\}\,.\quad
(T\ll M)
\label{3.4}
\end{equation}
}
Further if we assume $Ma\ll 1$, we immediately recognize that the
contribution of all the excited pyrgon states except for $\ell=0$ is
suppressed by Boltzmann factor because of their large masses.
Therefore we obtain the expression of $\rho$ when $T\ll M \ll 1/a$:
\begin{equation}
\rho\sim\left(\frac{M T_c}{2\pi}\right)^{d/2}\zeta(d/2)\,.
\quad (Ta\ll Ma \ll 1)
\label{3.5}
\end{equation}

On the other hand, if we suppose $Ma\gg 1$, (\ref{3.4}) becomes
\begin{eqnarray}
\rho&\sim&\sum_\ell d_\ell \left(\frac{T_c}{2\pi}\right)^{d/2}
(\sqrt{\omega_\ell^2+M^2})^{d/2}
\sum_{n=1}^\infty
\frac{1}{n^{d/2}}
\exp\left(-\frac{n\beta}{2M}\omega_\ell^2\right)\,.\nonumber \\
&&(T\ll M\mbox{ and }Ma\gg 1)
\label{3.6}
\end{eqnarray}
For this case, the approximation scheme is divided into two ways
according to he magnitude of $(MTa^2)$.

First, suppose $TMa^2\ll 1$. This condition is equivalent to
$1/Ta\gg Ma\gg 1$. Again highly massive modes are suppressed, then only
the contribution of $\ell=0$ survives. We obtain
\begin{equation}
\rho\sim\left(\frac{MT_c}{2\pi}\right)^{d/2}\zeta(d/2)\,.\quad
 (T\ll M \mbox{ and }TMa^2\ll 1)
\end{equation}

 Next, suppose $TMa^2\gg 1$. For this case, we can use the formula
\begin{equation}
\sum_\ell d_\ell \exp(-x\ell(\ell+N-1))\stackrel{x\ll 1}{\sim}
\frac{\Gamma(N/2)}{\Gamma(N)}x^{-N/2}\,.
\label{3.8}
\end{equation}
By use of this, we have
\begin{equation}
\rho\sim\left(\frac{MT_c}{2\pi}\right)^{(d+N)/2}\zeta((d+N)/2)V_N\,.\quad
 (T\ll M\mbox{ and }TMa^2\gg 1)
\label{3.9}
\end{equation}
where $V_N=(2\pi^{(N+1)/2}/\Gamma((N+1)/2))a^N$. The next
leading term will be small by the
factor $T/M$ or $1/TMa^2$ compared with the
leading (\ref{3.9}).

At high temperature $T\gg M$, there are both non-relativistic and
relativistic states since the system contains unlimited massive
states. A typical example is the case that only the $\ell=0$ state can
be relativistic, and others are nonrelativistic. Then the condition is
given by $T\gg M$ and $T\ll 1/a$. Whereas only $\ell=0$ term contributes
to $\rho$, for this time we use another asymptotic form of $K_\nu(z)$
\begin{equation}
K_\nu(z)\stackrel{z\ll
1}{\longrightarrow}\frac{2^{\nu-1}\Gamma(\nu)}{z^\nu}\,.
\label{3.10}
\end{equation}
Then we obtain
\begin{equation}
\rho\sim\frac{2MT_c^{d-1}}{\pi^{(d+1)/2}}
\Gamma\left(\frac{d+1}{2}\right)\zeta(d-1)\,.
\quad(1/a\ll M\ll T)
\label{3.11}
\end{equation}

Finally, we consider an extreme case, i.e., $T\gg M$ and $T\gg 1/a$. In
this case, high-temperature expansion is admitted,\cite{13} and it leads
to the following expression using (\ref{3.8}) and (\ref{3.10}):
\begin{equation}
\rho\sim\frac{2MT_c^{d+N-1}}{\pi^{(d+N+1)/2}}
\Gamma\left(\frac{d+N+1}{2}\right)\zeta(d+N-1)V_N\,.
\quad(T\gg M\mbox{ and }T\gg 1/a) 
\label{3.12}
\end{equation}

To summarize: we can obtain several approximate form of (\ref{3.1}) as
follows:
{\small
\begin{eqnarray}
\mbox{If } T\ll M\mbox{ and }T\ll 1/a,\mbox{ then}&&
\rho\sim\left(\frac{MT_c}{2\pi}\right)^{d/2}\zeta(d/2)\,;\nonumber \\
1\ll Ma\ll \frac{1}{Ta},&&
\rho\sim\left(\frac{MT_c}{2\pi}\right)^{(d+N)/2}\zeta((d+N)/2)V_N
\,;\nonumber
\\
M\ll T\ll 1/a,&&\rho\sim\frac{2MT_c^{d-1}}{\pi^{(d+1)/2}}
\Gamma\left(\frac{d+1}{2}\right)\zeta(d-1)\,;\nonumber \\
T\gg M\mbox{ and }T\gg 1/a,&&\rho\sim\frac{2MT_c^{d+N-1}}{\pi^{(d+N+1)/2}}
\Gamma\left(\frac{d+N+1}{2}\right)\zeta(d+N-1)V_N\,.\nonumber \\
\end{eqnarray}
}
\section{Effect of Kaluza-Klein charge asymmetry}
So far we have paid attention to the condensation of a charged boson.
Here the ``charge'' implies the one in the higher-dimensional theory. h
this section, we consider the Kaluza-Klein charge asymmetry. For
simplicity, we restrict ourselves to studying the abelian Kaluza-Klein
model.

First of all, we show the thermodynamic potential $\Omega$ with $\mu=0$
for a neutral scalar field in the geometry $R^d \times S^1$.
Note if $\mu=0$, we can call it ``free energy''. The
thermodynamic potential $\Omega$ is
\begin{eqnarray}
\Omega&=&-\frac{V_d}{(4\pi)^{(d+1)/2}}\int_0^\infty dt\,
t^{-(d+1)/2-1}\nonumber\\
&&\times\sum_{n=1}^\infty\sum_{\ell=-\infty}^\infty\exp
\left\{-\left(M^2+\frac{\ell^2}{a^2}\right)t
-\frac{\beta^2n^2}{4t}\right\}\,,
\label{4.1}
\end{eqnarray}
where $a$ is the radius of $S^1$.

On the other hand, the pyrgons with $(mass)^2=M^2+\ell^2/a^2$ have
``Kaluza-Klein charge'' which is proportional to $\ell/a$. By the
analogy of the results found in the preceding sections, we introduce
the chemical potential as follows:
\begin{eqnarray}
\Omega&=&-\frac{V_d}{(4\pi)^{(d+1)/2}}\int_0^\infty dt\,
t^{-(d+1)/2-1}\nonumber\\
&&\times\sum_{n=1}^\infty\sum_{\ell=-\infty}^\infty\cosh(n\beta\mu_\ell)\exp
\left\{-\left(M^2+\frac{\ell^2}{a^2}\right)t
-\frac{\beta^2n^2}{4t}\right\}\,,
\label{4.2}
\end{eqnarray}
where the relation $\mu_\ell=\ell\mu$ reflects the assumption of the
chemical equilibrium among pyrgon states.

The particle number can be derived from $\Omega$:
\begin{eqnarray}
\rho&=&-\frac{1}{V_d}\frac{\partial\Omega}{\partial\mu}=
\frac{1}{(4\pi)^{(d+1)2}}T^d\sum_{n=1}^\infty\sum_{\ell=1}^\infty
2n\ell\sinh(n\ell\mu\beta)\nonumber
\\ &&\times
2\left(\frac{2\beta\sqrt{M^2+\ell^2/a^2}}{n}\right)^{(d+1)/2}K_{(d+1)/2}
\left(\beta\sqrt{M^2+\frac{\ell^2}{a^2}}n\right)\,.
\label{4.3}
\end{eqnarray}

Incidentally, we can express $\rho$ in a familiar form:
\begin{equation}
\rho=\sum_{\ell=1}^\infty\int\frac{d^d{\bf k}}{(2\pi)^d}
\left[\frac{1}{\exp\{\beta(\sqrt{M^2+\ell^2/a^2}-\ell\mu)\}-1}
-(\mu\rightarrow-\mu)\right]\,.
\label{4.4}
\end{equation}
If the value of $\rho$ increases without any limit when we change the
value of $\mu$, We can say that there exists no condensation in this
system. In the above case, we can see when $\mu>1/a$ we cannot give
physical meaning to $\rho$ or $\Omega$. Thus, we need to investigate the
behavior of $\rho$ when $\mu$ increases toward $1/a$.

In order to find whether the summation will converge or not, we look
into the behavior of the functions at large $\ell$ or $n$.

Since the asymptotic form of the modified Bessel function is
\begin{equation}
K_\nu(z)\stackrel{z\rightarrow\infty}{\sim} 
\sqrt{\frac{\pi}{2z}}e^{-z}\,,
\label{4.5}
\end{equation}
thus, the series (\ref{4.2}) becomes asymptotically as follows:
\begin{eqnarray}
&&\sum\sum\frac{V_d}{(4\pi)^{(d+1)/2}}T^d 2\sqrt{\pi}\ell
\left(2\beta\sqrt{M^2+\frac{\ell^2}{a^2}}\right)^{d/2}\nonumber\\
&&\times n^{-(d-1)/2}
\exp\left\{\beta
n\left(\ell\mu-\sqrt{M^2+\frac{\ell^2}{a^2}}\right)\right\}\,.
\label{4.6}
\end{eqnarray}
The prefactor of the exponential in the series turns out to be an
increasing function in $\ell$. Thus the summation on $\ell$ will be
finite provided that the exponential function in (\ref{4.6}) suppresses
the large-$\ell$ terms. However, a closer looking into the exponential
reveals that if $\mu=1/a$ the exponential is not damped at large
$\ell$. Accordingly, the sum of the series diverges when
$\mu\rightarrow 1/a$. After all, we conclude that the Bose-Einstein
condensation does not occur in this system.

We can arrive at this conclusion by taking the following facts
into consideration; first, there are infinite number of states which
come from the compactification; second, at the same temperature, the
particle system of the larger mass can include the larger number of
particles (e.g., see (\ref{2.6})).

Of course, the assumption of the chemical equilibrium among the
pyrgon states plays an essential role in this calculation. The
physical argument for this will be given in the next section.

\section{Discussion}
In this paper, we have investigated the Bose-Einstein condensations in
higher dimensions and in the Kaluza-Klein space.

We treated here only the free scalar fields and discussed an abelian
charge or a particle number. It might be most interesting to check if
these results are still valid when interaction is included. It is
obvious that, in order to deal with the decay of the particles and
non-equilibrium process, we should reformulate our method in deriving
the thermodynamic quantities so as to take into account the various
interactions and time-dependence of the background geometry. In
particular, the argument for the Kaluza-Klein charge in the previous
section will not be valid if we take into account the interaction in
an expanding universe where the coupling ``constant'' itself varies
when the radius of the extra space changes. Furthermore, the form of
the interactions restricts the transition between pyrgon states, and
the naive assumption of the chemical equilibrium is not suitable.
This must be taken into account when we make the generalization to
non-abelian Kaluza-Klein theories.

Further, the curvature and/or the dynamical behavior of the
background geometry will induce the interactions which violate the
conservation of the Kaluza-Klein charge; this possibility is pointed
out by Orito and Yoshimura.\cite{14} As well, the condition of
equilibrium is to be re-considered.

To investigate the cosmological aspects of the behavior of matters,
it is appropriate to use the so-called ``real-time formalism'' or TFD
(Thermofield Dynamics)\cite{15} arranged more axiomatically, which has
been developed recently. Particularly, irreversible and nonequilibrium
processes in the universe attract much attention.\cite{16} It is
interesting to study the decay of pyrgons which gives rise to the
production of entropy in the universe.\cite{12,17}  Also the evolution
of the scale of the compact space and its oscillation around the
constant value will be affected by the irreversible process such as
the particle production.\cite{18}

We are also interested in the study of the behavior of gauge fields
in higher-dimensional theory at finite temperature. Superstring
theories imply the presence of the gauge field as ``primary
field''.\cite{5} The finite temperature effect on the break-down of
primary gauge symmetries as well as compactifications is expected to
be important in the early universe. It is necessary to investigate the
symmetry breaking by the non-trivial Wilson line \cite{19} in the
evolving multidimensional universe.

\section*{Acknowledgements}
The author would like to thank M. Hosoda for useful discussion and
reading of this manuscript.


\end{document}